\documentclass[fleqn,10pt,a4paper]{article}
\usepackage{amsmath,amssymb,bm}
\usepackage{cite,epsfig,color,graphicx}

\title{Non-perturbative aspects of particle acceleration in non-linear electrodynamics}

\author{David A. Burton\thanks{Department of Physics, Lancaster University, Lancaster, UK and Cockcroft Institute, Daresbury, UK.}\and
Stephen P. Flood\footnotemark[1]\and
Haibao Wen\footnotemark[1]\,\,\thanks{Current address: Department of Earth Science \& Engineering, Imperial College London, UK.}}

\begin{document}
\maketitle
\begin{abstract}
We undertake an investigation of particle acceleration in the context of non-linear electrodynamics. We deduce the maximum energy that an electron can gain in a non-linear density wave in a magnetised plasma, and we show that an electron can `surf' a sufficiently intense Born-Infeld electromagnetic plane wave and be strongly accelerated by the wave. The first result is valid for a large class of physically reasonable modifications of the linear Maxwell equations, whilst the second result exploits the special mathematical structure of Born-Infeld theory.   
\end{abstract}

\section{Introduction}
The implications of theories that couple the electromagnetic field to itself have been an enduring source of interest to particle theorists for decades, and recent developments in ultra-high intensity lasers have led to a surge of interest in relativistic non-linear electrodynamics by the wider community. It is expected that facilities such as ELI~\cite{eli} will permit investigation of laser-matter interactions at intensities where non-linear perturbative effects mediated by virtual electron-positron pairs will be evident~\cite{dipiazza:2012}, and numerous studies of the implications of the Euler-Heisenberg Lagrangian have been undertaken in this context. In the longer term, it is anticipated that vacuum pair production, a fundamentally non-perturbative QED process, will become accessible in the laboratory (for a recent review of the attendant theoretical issues see Ref.~\cite{dunne:2014}). In addition to QED processes, attention has also been paid in recent years to tests of axion electrodynamics using ultra-high intensity lasers~\cite{dipiazza:2013}.

Considerable progress has been made during recent years in the exploitation of large-amplitude plasma waves for particle acceleration. Such schemes are particularly attractive in the laboratory because the electric fields in a plasma wave can be several orders of magnitude greater than those sustainable in standard radio-frequency accelerator cavities. The most prevalent schemes realised thus far employ the strong fields in the wake behind an intense laser pulse propagating through the plasma~\cite{schlenvoigt:2008}, although electron-driven wakefields have also been exploited for electron acceleration~\cite{blumenfeld:2007}. Furthermore, recent developments have focussed on proton-driven wakefields as a paradigm for efficiently accelerating leptons to TeV energies~\cite{assman:2014}. In the astrophysical context, plasma waves were recently invoked to explain the emission of energetic electrons from within the interiors of pulsars~\cite{diver:2010}; such electrons are necessary for the formation of the electron-positron plasma populating a pulsar's magnetosphere. The magnetic fields found in neutron stars are typically $\sim 10^8\,{\rm T}$, whilst those in magnetars may be two orders of magnitude higher, and non-linear effects due to QED are expected to be significant in such environments~\cite{lai:2003}. One may speculate that non-Standard Model couplings also play an important role.

The Born-Infeld Lagrangian is probably the most famous theory of non-linear electrodynamics whose motivation lies outside the Standard Model. It first appeared in the 1930s as a classical model of the electron with finite self-energy~\cite{born:1934}, resurfaced during the mid-1980s as an effective action in string theory~\cite{fradkin:1985} and its notoriety was finally cemented during subsequent years in the context of $D$-branes~\cite{callan:1998}. The Born-Infeld Lagrangian has a privileged mathematical status within the family of Lagrangians that depend only on the electromagnetic field tensor, the dual of the electromagnetic field tensor and the spacetime metric tensor (but not their derivatives). It is the only such regular generalization of the vacuum Maxwell Lagrangian whose field equations exhibit an absence of birefringence and shocks~\cite{plebanski:1968, boillat:1970, gibbons:2001}. 

Such considerations led to a recent study~\cite{burton_trines_walton_wen:2011} of the properties of non-linear density waves in a Born-Infeld plasma and their implications for particle acceleration. An approximation to the maximum energy gain of an electron trapped and accelerated by the wave was obtained, but the result was independent of the Born-Infeld coupling parameter. Section~\ref{sec:non-lin_plasma_wave} below revisits this topic from a general perspective with an investigation of particle acceleration in a non-linear density wave propagating along the ambient field lines of a magnetised plasma with arbitrary electromagnetic self-couplings. We show that the non-appearance of the Born-Infeld coupling parameter in the estimate of the maximum energy gain of the trapped electron in Ref.~\cite{burton_trines_walton_wen:2011} has nothing to do with the exceptional properties of the Born-Infeld Lagrangian. Indeed, we demonstrate that an exact calculation of the maximum energy gain yields the same result for any physically reasonable Lagrangian that is an algebraic expression of the two fundamental invariants of the electromagnetic field, and this result is independent of the strength of the ambient magnetic field.  

Unlike many other non-linear theories, the source-free Born-Infeld field equations possess numerous non-trivial exact solutions which are indispensible for exploring non-perturbative aspects of the theory. In particular, an exact solution describing an electromagnetic pulse immersed in a uniform magnetic field is known~\cite{aiello:2007} and we suggest that this may have implications for vacuum laser acceleration in future facilities, such as ELI~\cite{eli}. Section~\ref{sec:non-lin_vacuum_wave} shows that an electron interacting with a non-linear Born-Infeld electromagnetic plane wave can be uniformly accelerated to arbitrarily high energies. This novel result is non-perturbative and has no analogue in linear Maxwell electromagnetics.   

Heaviside-Lorentz units are used throughout the following with $c=1$ and, to avoid an unnecessary plethora of indices, intrinsic geometrical notation is used extensively. Further details of the notation and conventions used here may be found in Ref.~\cite{burton:2003}. 
\section{Particle acceleration in a strongly magnetised plasma}
%%%%
\label{sec:non-lin_plasma_wave}
Much attention has been devoted to uncovering the behaviour of particles trapped in an electron density wave driven by an intense laser pulse, or particle bunch, propagating through a plasma. Although fully 3-dimensional configurations (the `bubble regime') now pervade such studies~\cite{lu:2006, shvets:2013}, the original laser-wakefield accelerator concept was formulated using 1-dimensional considerations~\cite{tajima:1979} and approaches that employ non-linear plane waves remain useful for providing estimates. Furthermore, one can argue that plane waves are appropriate for modelling particle acceleration in neutron star crusts. The strong magnetic field within a neutron star polarizes the iron outer crust of the neutron star and leads to a highly anisotropic conductivity~\cite{lai:2001}. Electron density waves excited within the magnetic flux tubes are essentially free to propagate along the magnetic field, but their motion transverse to the field is greatly restricted; hence, to a first approximation, it is sufficient to only consider motion along the magnetic field lines~\cite{diver:2010}.

The purpose of this section is to show that the maximum gain in energy of an electron in a non-linear plane wave in a magnetised plasma is invariant within a large class of theories of electromagnetism. The plasma is modelled as a superposition of two charged pressureless perfect fluids, where one fluid describes mobile electrons and the other describes the charge carriers of a neutralizing background medium. It is assumed that the spacetime curvature is negligible and the worldlines of the charge carriers of the neutralizing background are timelike geodesics. Hence, we adopt the Minkowski metric
\begin{equation}
g = - dt \otimes dt + dx \otimes dx + dy \otimes dy + dz \otimes dz
\end{equation}
and choose the $4$-velocity $V_0$ of the neutralizing background to be $V_0 = \partial_t$.

The latter approximations are reasonable if the motion of the background is negligible over the timescales of interest. This is certainly the case if the background consists of ions whose charge-to-mass ratio is approximately three orders of magnitude lower than that of the mobile electrons (as in a wakefield accelerator), or if the background is the polarised outer crust of a neutron star and non-inertial terms (due to the rotation of the neutron star) in the metric components can be ignored.

The worldlines of the mobile electrons are trajectories of the unit normalised future-pointing timelike $4$-vector field $V$. It is assumed that the motion of the electrons is parallel to a constant ambient magnetic field and their $4$-acceleration $\nabla_V V$ satisfies
\begin{equation}
\label{lorentz}
\nabla_V \widetilde{V} = \frac{q}{m}\iota_V F
\end{equation}
with
\begin{equation}
\label{V_norm}
g(V,V)=-1
\end{equation}
where $q \iota_V F$ is the Lorentz $4$-force acting on the mobile electrons, $\iota_V$ is the interior product with respect to $V$, $-q=e$ is the elementary charge, $m$ is
the electron rest mass, $F$ is the electromagnetic $2$-form and $\nabla$ is the Levi-Civita connection. The $1$-form $\widetilde{V}$ is the metric dual of the vector field $V$ i.e. the $1$-form $\widetilde{V}$ satisfies $\widetilde{V}(U)=g(V,U)$ for all vector fields $U$. Although the electromagnetic self-force (radiation reaction) on the electrons plays a significant role in strong fields~\cite{burton:2014}, little is known about the self-force outside of the context of classical linear electromagnetism or perturbative QED. The emphasis of the following is on maintaining generality in the electromagnetic sector, and so the Lorentz force is adopted in the absence of a theory of radiation reaction in the general context.

The background fluid is described by the electric $4$-current $e n_0 V_0$ where the constant $n_0$ has the physical dimensions of a number density. In the case of a plasma, $n_0/Z$ is the proper number density of the ions where $Z$ is the ionisation multiplicity. The electromagnetic field equations may be written covariantly as
\begin{equation}
\label{maxwell}
dF = 0,\qquad d\star G = - q n\star\widetilde{V} + q n_0 \star\widetilde{V_0}
\end{equation}
where $n$ is the proper number density of the electron fluid and the Hodge map $\star$ is induced from the $4$-form $\star 1$ given as
\begin{equation}
\star 1 = dt \wedge dx \wedge dy \wedge dz.
\end{equation}

A constitutive relation specifying the excitation $2$-form $G$ in terms of the $2$-form $F$ must be given to close the system of field equations (\ref{lorentz}), (\ref{V_norm}), (\ref{maxwell}). Equations (\ref{lorentz}), (\ref{V_norm}), (\ref{maxwell}) may be derived~\cite{burton_trines_walton_wen:2011} from a Lagrangian containing a $0$-form-valued function ${\cal L}_{\rm EM}$ of the invariants $X$ and $Y$ defined as
\begin{equation}
\label{invariants}
X = \star(F\wedge \star F),\qquad Y = \star(F\wedge F),
\end{equation}
where
\begin{equation}
\label{starG}
\star G = 2\bigg(\frac{\partial{\cal L}_{\rm EM}}{\partial X}\star F + \frac{\partial{\cal L}_{\rm EM}}{\partial Y} F\bigg)
\end{equation}
and the choice ${\cal L}_{\rm EM} = X/2$ yields classical linear Maxwell theory.

For present purposes it is advantageous to replace (\ref{lorentz}) with 
\begin{equation}
\label{Killing_stress_balance}
d\tau_K = j_0\wedge\iota_K F 
\end{equation}
where $j_0 = q n_0 \star\widetilde{V_0}$ and $K$ is a Killing vector.  The stress(-energy-momentum) $3$-form $\tau_K$ is the sum of contributions from the electromagnetic field and the electron fluid:
\begin{equation}
\label{Killing_stress_forms}
\tau_K = \underset{\text{EM field}}{\underbrace{\iota_K F \wedge \star G + {\cal L}_{\rm EM}\star\widetilde{K}}} + \underset{\text{electron fluid}}{\underbrace{mn\,g(K,V)\star\widetilde{V}}}.
\end{equation}
Equation (\ref{Killing_stress_balance}) expresses the local balance of energy, momentum or angular momentum when $K$ is chosen appropriately. For example, if $K$ is a unit timelike Killing vector then (\ref{Killing_stress_balance}) connects the rate of change of the total energy density of the electromagnetic field and the electron fluid with the rate of work done per unit volume by the Lorentz force on the background fluid. Balance laws associated with linear or angular momentum are revealed when $K$ is chosen to be a generator of translations or rotations, respectively. It can be shown that the two systems of equations (\ref{lorentz}), (\ref{V_norm}), (\ref{maxwell}) and (\ref{Killing_stress_balance}), (\ref{V_norm}), (\ref{maxwell}) are equivalent.

The above model is applicable to a wave in the electron fluid density propagating along the lines of an ambient magnetic field whose curvature and evolution can be neglected. In this case, the electromagnetic field $F$ is a superposition of an ambient homogenous static magnetic field $(0, 0, B)$ and the electric field $(0,0,E(\zeta))$ driven by the electron fluid:
\begin{equation}
\label{electric_wave_ansatz}
F = E(\zeta)\, dt\wedge dz - B\, dx\wedge dy
\end{equation}
where $\zeta = z - v t$ is the phase of the density wave and the phase velocity $v$ is a constant satisfying $0<v<1$. Although no generality is lost by requiring $v>0$, the choice $|v|<1$ is a critical ingredient in the wakefield accelerator paradigm since particles must have the opportunity to be trapped in the wave and accelerated.

It has long been known that non-linearities arising purely from the matter content lead to an upper bound on the amplitude of the electric field of a steady density wave whose wave $4$-vector is spacelike~\cite{akhiezer:1956, dawson:1959}. The maximum amplitude, known as the `wave-breaking limit' by the plasma accelerator community, is an important parameter in the wakefield accelerator paradigm and is sensitive to the details of the plasma model~\cite{burton:2010, burton_trines_walton_wen:2011}. Although the magnetic field considered here is aligned with the velocity of the electron fluid and cannot directly influence their motion, it may influence their motion indirectly through electromagnetic self-coupling in the excitation $2$-form (\ref{starG}) and thereby affect the wave-breaking limit.

It is convenient to analyse the field equations (\ref{Killing_stress_balance}), (\ref{V_norm}), (\ref{maxwell}) using the pair $\{e^1,e^2\}$:
\begin{equation}
e^1= vdz - dt, \qquad e^2=dz - vdt = d\zeta
\end{equation}
where the orthonormal coframe $\{\gamma e^1,\gamma e^2,
dx, dy\}$, with $\gamma=1/\sqrt{1-v^2}$, is adapted to observers moving at velocity $v$ along
$z$ in the rest frame of the neutralizing background medium (i.e observers at rest in the `wave frame'). We seek a $4$-velocity field $V$ of the form
\begin{equation}
\label{V_ansatz}
\widetilde{V} = \mu(\zeta) e^1 + \psi(\zeta) e^2
\end{equation} 
where, using $g(V,V) = -1$, the component $\psi$ is
\begin{equation}
\label{negative_psi}
\psi = -\sqrt{\mu^2-\gamma^2}
\end{equation}
with the sign of $\psi$ chosen to ensure that the velocity $\gamma
e^2 (V)$ of the plasma electrons in the wave frame is
non-positive. Hence, the speed of the plasma electrons in the frame of the neutralizing medium is less than the phase speed $v$ of the wave.

The components of $G$ in the basis $\{dt, dx, dy, dz\}$ depend only on $\zeta$ and it follows that $d\zeta\wedge d\star G=0$. Hence, (\ref{maxwell}) yields the electron proper number density $n$ as a function of $\mu$,
\begin{equation}
\label{proper_num_density}
n = \frac{n_0 v \gamma^2}{\sqrt{\mu^2-\gamma^2}}, 
\end{equation}
allowing $n$ to be eliminated from the stress $3$-form $\tau_K$.

Analysis of the balance law (\ref{Killing_stress_balance}) proceeds by choosing the $1$-form $\widetilde{K}$ to be each member of the basis $\{e^1, e^2,
dx, dy\}$ in turn. Both sides of (\ref{Killing_stress_balance}) vanish when $\widetilde{K}\in\{dx, dy\}$ (the momentum transferred between the electromagnetic field and electron fluid is along $z$ only) whereas
\begin{equation}
\label{stress_1}
\tau_K \simeq m n_0 v\mu\, e^1\wedge dx \wedge dy 
\end{equation}
for $K = \widetilde{e^1}$, with $\simeq$ indicating equality modulo closed forms and where $n$ has been eliminated using (\ref{proper_num_density}). Insertion of (\ref{stress_1}) into (\ref{Killing_stress_balance}) leads to
\begin{equation}
\label{E_field}
E = \frac{m}{q}\frac{1}{\gamma^2}\frac{d\mu}{d\zeta}. 
\end{equation}
Equations (\ref{invariants}), (\ref{electric_wave_ansatz}) lead to $X=E^2-B^2$, $Y=2EB$, which with $K = \widetilde{e^2}$ yields
\begin{align}
\label{stress_2}
\tau_K \simeq &\bigg[-2 E^2 \frac{\partial{\cal L}_{\rm EM}}{\partial X} - 2 E B \frac{\partial{\cal L}_{\rm EM}}{\partial Y} + {\cal L}_{\rm EM}
+ m n_0 v\sqrt{\mu^2-\gamma^2} \bigg]\, e^1\wedge dx \wedge dy,
\end{align}
and
\begin{equation}
\label{ion_work}
j_0 \wedge \iota_K F = m n_0 \frac{d\mu}{d\zeta}\,e^2\wedge e^1\wedge dx \wedge dy 
\end{equation}
where (\ref{E_field}) has been used. 

Hence, equations (\ref{Killing_stress_balance}), (\ref{stress_2}), (\ref{ion_work}) give
\begin{align}
\label{energy_balance}
\frac{d}{d\zeta}\bigg[2 E^2 \frac{\partial{\cal L}_{\rm EM}}{\partial X} + 2 E B \frac{\partial{\cal L}_{\rm EM}}{\partial Y} - {\cal L}_{\rm EM}
- m n_0 (v\sqrt{\mu^2-\gamma^2} - \mu)\bigg]  = 0.
\end{align}

The periodic solutions to (\ref{energy_balance}) for $\mu$ have certain properties. Clearly $\mu \ge \gamma$ due to the square root in (\ref{energy_balance}) and the periodic solution with the largest amplitude satisfies $\mu(\zeta_{\rm I})=\gamma$ where $\zeta=\zeta_{\rm I}$ is a zero of $d\mu/d\zeta$ (and hence, using (\ref{E_field}), $E(\zeta_{\rm I})=0$). Thus, the largest amplitude solution to (\ref{energy_balance}) satisfies
\begin{align}
\label{largest_amp_sol}
2 E^2 \frac{\partial{\cal L}_{\rm EM}}{\partial X} + 2 E B \frac{\partial{\cal L}_{\rm EM}}{\partial Y} - {\cal L}_{\rm EM}
- m n_0 (v\sqrt{\mu^2-\gamma^2} - \mu) = m n_0 \gamma - {\cal L}_{\rm EM}|_{E=0}.
\end{align}
The maximum value $\mu_{\rm max}$ of $\mu$ may be determined by evaluating (\ref{largest_amp_sol}) at the turning point $\zeta=\zeta_{\rm II}$ of $\mu$ immediately after the turning point $\zeta=\zeta_{\rm I}$. Using (\ref{E_field}) and $d\mu/d\zeta|_{\zeta_{\rm II}} = 0$ in (\ref{largest_amp_sol}) yields
\begin{equation}
\mu_{\rm max} = \mu(\zeta_{\rm II}) = \gamma^3(1+v^2). 
\end{equation}

Thus far, we have only considered the mobile electrons that form the wave (i.e. those described by the electron fluid). We now turn to the behaviour of electrons captured by the density wave. However, electrons trapped in the density wave do not, in general, follow the worldlines of the electron fluid (since, in contrast to the outcome of (\ref{negative_psi}), they may propagate faster than the wave) and, in general, elucidating the impact of the captured electrons requires intensive numerical computation. However, if the population of captured electrons is sufficiently small then their back-reaction on the wave can be neglected. In this case, the motion of the captured electrons is dictated by the total electromagnetic field of the electron fluid, the neutralizing background and the background magnetic field only; hence, the captured electrons are modelled as test particles.

The electric field is static in the wave frame and it follows that the quantity
\begin{equation}
\label{Delta_mu}
\Delta\mu \equiv \mu(\zeta_{\rm II})-\mu(\zeta_{\rm I})= 2\gamma^3 v^2
\end{equation}
is proportional to potential difference and therefore proportional to the increase in the energy of a test electron that moves from $\zeta=\zeta_{\rm I}$ to $\zeta=\zeta_{\rm II}$. The coefficient of proportionality is straightforward to obtain using the following elegant argument.

The test electron's worldline $C$ satisfies the Lorentz equation
\begin{equation}
\label{test_lorentz}
m\nabla_{\dot{C}}\widetilde{\dot{C}} = q\iota_{\dot{C}}F
\end{equation}
where the electron's $4$-velocity $\dot{C}$ satisfies $g(\dot{C},\dot{C})=-1$. The unit timelike Killing vector $K = \gamma(\partial_t + v\partial_z)$ satisfies $\nabla_{\dot{C}} K = 0$ and it may be shown that $d[g(\dot{C},K)]/d\tau = (\nabla_{\dot{C}} \widetilde{\dot{C}})(K)$ where $\tau$ is the electron's proper time. Hence $d[- m g(\dot{C},K)]/d\tau = q \iota_{\dot{C}} \iota_K F$ where $-m g(\dot{C},K)$ is the energy of the electron in the wave frame, and it follows that the change in energy of the electron over the interval $[\tau_{\rm I}, \tau_{\rm II}]$ is $q \int^{\tau_{\rm II}}_{\tau_{\rm I}} \iota_{\dot{C}}\iota_K F\,d\tau$. The previous integral may be written in the covariant and parameterisation-independent form $q\int_C \iota_K F$ and the change in energy $\Delta E_K$,
\begin{equation}
\Delta E_K = - m g(\dot{C},K)\big|^{\tau_{\rm II}}_{\tau_{\rm I}},
\end{equation}
of the test electron in the wave frame follows immediately: 
\begin{equation}
\label{wave_frame_energy_gain}
\Delta E_K = q\int_C \iota_K F = \int\limits^{\zeta_{\rm II}}_{\zeta_{\rm I}} \frac{m}{\gamma} \frac{d\mu}{d\zeta}\,d\zeta = \frac{m}{\gamma}\Delta\mu = 2m\gamma^2 v^2.
\end{equation}
Previous estimates of the maximum energy gain~\cite{diver:2010, burton_trines_walton_wen:2011} were obtained using quantities, such as the maximum electric field of the density wave, that require ${\cal L}_{\rm EM}(X,Y)$ to be specified explicitly. In the above, we bypassed such details and obtained a lower bound on the electron's maximum energy gain independent of ${\cal L}_{\rm EM}(X,Y)$. Although the profile of the density wave depends on the details of ${\cal L}_{\rm EM}(X,Y)$, the potential difference between two adjacent nodes of the maximum amplitude electric field is independent of ${\cal L}_{\rm EM}(X,Y)$. If the maximum amplitude density wave is weaker than its counterpart in classical Maxwell electromagnetism, then the period of the wave must be longer to ensure that (\ref{wave_frame_energy_gain}) holds. The above result is immutable and explains why the estimate of the maximum energy gain in the unmagnetised Born-Infeld plasma wave given in Ref.~\cite{burton_trines_walton_wen:2011} is independent of the Born-Infeld parameter.

Unlike the results in the wave frame, the energy gain of the test electron in the frame of the neutralizing background depends on the electron's initial conditions. As with the electron fluid, we will assume that the test electron's motion is along the magnetic field lines only.

The electron's $4$-velocity $\dot{C}$ may be written as $\dot{C} = \gamma_u(K + u L)$ where $K=\gamma(\partial_t + v \partial_z)$, $L = \gamma(\partial_z + v \partial_t)$, $\gamma_u = 1/\sqrt{1-u^2}$ and $(0, 0, u)$ is the $3$-velocity of the electron in the wave frame. Thus, the change in energy $\Delta E_{\partial_t}$ of the electron in the frame of the neutralizing background is
\begin{equation}
\label{lab_frame_energy_gain}
\Delta E_{\partial_t} = -m g(\dot{C},\partial_t)\big|^{\tau_{\rm II}}_{\tau_{\rm I}} = m\gamma\big[\gamma_{\rm II}(1 + u_{\rm II} v) - \gamma_{\rm I}(1 + u_{\rm I} v)\big]
\end{equation}
where $u_{\rm I}$, $u_{\rm II}$ and $\gamma_{\rm I}$, $\gamma_{\rm II}$ are the values of $u$ and $\gamma_u$, respectively, at the points $\rm{I}$, $\rm{II}$ on $C$. Since $-g(\dot{C},K)\big|^{\tau_{\rm II}}_{\tau_{\rm I}} = \gamma_{\rm II} - \gamma_{\rm I}$ follows trivially from the definition of $K$ we find 
\begin{equation}
\label{wave_frame_gamma_gain}
\gamma_{\rm II} - \gamma_{\rm I} = 2 \gamma^2 v^2
\end{equation} 
using (\ref{wave_frame_energy_gain}), and (\ref{lab_frame_energy_gain}), (\ref{wave_frame_gamma_gain}) can be used to express $\Delta E_{\partial_t}$ in terms of $\gamma_{\rm I}$ and $v$. The electron begins at a node of the electric field and we choose $u_{\rm I}>0$; it can then be shown that $d\Delta E_{\partial_t}/d\gamma_{\rm I} < 0$ and it follows that the largest value of $\Delta E_{\partial_t}$, for fixed $v$, is $\lim_{\gamma_{\rm I} \rightarrow 1}\Delta E_{\partial_t} = 4 m(\gamma^3 - \gamma)$. An electron that starts at a node of the electric field with speed $u=0^+$ in the wave frame, and reaches the adjacent node, gains energy $\Delta E_{\partial t} = 4 m(\gamma^3 - \gamma)$ in the frame of the neutralizing background medium. Thus, the maximum change in energy of an electron that begins at rest in the frame of the background is $4 m(\gamma^3 - \gamma) + m(\gamma - 1) = m (4\gamma^3 - 3\gamma - 1)$. Although the latter result was previously derived in the context of classical linear electromagnetism~\cite{esarey:1995}, using a very different approach to that presented here, our novel observation is that the same result holds for more general theories of electromagnetism.

Although the maximum energy gain does not explicitly depend on the details of the Lagrangian ${\cal L}_{\rm EM}(X,Y)$, it may implicitly depend on ${\cal L}_{\rm EM}(X,Y)$ since it depends on the phase speed $v$ of the density wave. In particular, if the density wave is driven by an ultra-strong laser pulse then $v$ is related to the structure and behaviour of the pulse which, in turn, are influenced by the details of ${\cal L}_{\rm EM}(X,Y)$ and the total electric $4$-current.
\section{Particle acceleration in a Born-Infeld plane wave}
%%%
\label{sec:non-lin_vacuum_wave} 
Several directions for generalizing the previous analysis are possible, and all involve an electromagnetic $2$-form $F$ whose structure is more general than (\ref{electric_wave_ansatz}). An option accessible to non-perturbative analysis is to focus on particle acceleration in regions where the total electric $4$-current can be neglected. Indeed, the on-going evolution of ultra-high-power laser technology has greatly increased the feasibility of directly accelerating matter in free space (so-called `vacuum laser acceleration') and the implications of an effective self-coupling of the electromagnetic field are of considerable interest in this context.

Before turning to non-linear electromagnetism, it is worth noting that plane wave solutions to the linear vacuum Maxwell equations are not normally consider to be useful for particle acceleration~\cite{heinzl:2009}.
In particular, suppose that a {\it pulsed} plane wave collides with an electron. After a sufficient time interval, the pulse must completely overtake the electron (regardless of the electron's initial conditions) because, in classical linear Maxwell theory, plane waves propagate rectilinearly at speed $c=1$ without dispersing.
However, it is common to choose an electric field whose profile is an antisymmetric function of the plane wave's phase when exploring the behaviour of an electron driven by an intense few-cycle laser pulse~\cite{harvey:2011, kravets:2013}, and therefore the integral of the electric field over the length of the pulse vanishes. It follows that there is no net change in the energy of the electron if radiation reaction is neglected.  

To be more precise, consider the $2$-form $F = {\cal E}(z - t)\,(dz - dt) \wedge dx$ where the once-differentiable function ${\cal E} : \mathbb{R} \rightarrow \mathbb{R}$ has compact support. The electric and magnetic fields of the pulse are $(-{\cal E},0,0)$ and $(0,-{\cal E},0)$ respectively. If the worldline $C : \tau \mapsto (t(\tau), x(\tau), y(\tau), z(\tau))$ of the electron satisfies the Lorentz equation $m\nabla_{\dot{C}}\widetilde{\dot{C}} = q\iota_{\dot{C}}F$ with $g(\dot{C},\dot{C})=-1$, and $\dot{C}$ is the particle's $4$-velocity with $\tau$ the electron's proper time,
then it is straightforward to show that $m \dot{t}(\tau_{\rm II}) = m \dot{t}(\tau_{\rm I})$ if $\int^{\phi_{\rm II}}_{\phi_{\rm I}}{\cal E}(\phi)\,d\phi = 0$ where $\phi = z - t$ and $\dot{t} = dt/d\tau$. It follows that the difference between the initial and final values of the electron's relativistic energy $m\dot{t}$ vanishes. Vacuum laser acceleration relies on using a {\it tightly-focussed} ultra-high-power laser pulse to directly accelerate electrons in free space~\cite{shao:2013}, and the effects of the pulse cannot be adequately captured by modelling it as a plane wave; it is vital to account for the non-trivial pointwise dependence of the electromagnetic field on $x, y$. However, there is no reason to conclude that plane waves are ineffective for particle acceleration if the effects of the quantum vacuum (of the Standard Model or otherwise) are manifest. Indeed, as we will now show, a non-perturbative analysis in this context yields new results that are inaccessible using perturbation theory.

The non-linear generalization of vacuum Maxwell electromagnetism introduced by Born and Infeld~\cite{born:1934} is a privileged theory. It is the only theory generated by a Lagrangian of the two electromagnetic invariants $X$, $Y$ that agrees with vacuum Maxwell theory in the weak-field regime and whose solutions do not exhibit birefringence and do not develop shocks~\cite{boillat:1970, plebanski:1968, gibbons:2001}. The Born-Infeld field equations also emerge from string/M theory~\cite{fradkin:1985, callan:1998}, and this has ignited modern interest~\cite{flood:2012, burton_trines_walton_wen:2011, dereli:2010, ferraro:2010a, munoz:2009, ferraro:2007, aiello:2007, denisov:2000} in it as an effective theory of electromagnetism in strong fields.

The source-free Born-Infeld field equations are
\begin{equation}
\label{BI_field_eqns}
dF = 0,\qquad d\star G_{\rm BI} = 0,
\end{equation} 
where the excitation $2$-form $G_{\rm BI} = 2(\partial_X{\cal L}_{\rm BI} F - \partial_Y{\cal L}_{\rm BI}\star F)$ is generated from the following $0$-form:
\begin{equation}
\label{definition_LBI}
{\cal L}_{\rm BI}(X,Y) = \frac{1}{\kappa^2}(1-\sqrt{1-\kappa^2 X - \kappa^4 Y^2/4}).
\end{equation}
The Born-Infeld parameter $\kappa$ controls the strength of the self-coupling of the electromagnetic field, and vacuum Maxwell theory is recovered in the limit $\kappa\rightarrow 0$. 

A substantial number of exact solutions to (\ref{BI_field_eqns}) have been discovered~\cite{aiello:2007, ferraro:2010b, ferraro:2013} despite the non-linear structure of Born-Infeld electromagnetism. In particular, the exact solution~\cite{aiello:2007}  
\begin{align}
\label{exact_wave_BI}
F = {\cal E}(z - v t)\,(dz - v dt) \wedge dx - B_x\,dy\wedge dz - B_y\,dz \wedge dx
- B_z\,dx \wedge dy + \chi {\cal E}(z - v t) dt \wedge dz
\end{align}
to (\ref{BI_field_eqns}) describes an electromagnetic plane wave propagating through an ambient uniform magnetic field $(B_x, B_y, B_z)$ where
\begin{equation}
\label{BI_wave_parameters}
\chi = \frac{\kappa^2 B_z B_x v}{1+\kappa^2 B_z^2},\qquad v=\sqrt{\frac{1+\kappa^2 B_z^2}{1+\kappa^2 B^2}}, 
\end{equation}
and $B = \sqrt{B_x^2 + B_y^2 + B_z^2}$. The wave propagates along the $z$-axis with phase velocity $v$ and its profile is encoded by the function ${\cal E} : \mathbb{R} \rightarrow \mathbb{R}$. The electric field measured by an inertial observer with $4$-velocity $\partial/\partial t$ is $(E_x=-v{\cal E}, E_y = 0, E_z = \chi {\cal E})$ and, in general, the electric field of a Born-Infeld electromagnetic wave has a non-zero longitudinal component in addition to the usual transverse component found in vacuum Maxwell theory. Remarkably, the smooth function ${\cal E}$ is essentially unconstrained; the only requirement is that it satisfies the bound 
\begin{equation}
1-\kappa^2 X - \kappa^4 Y^2/4 > 0
\label{XY_bound}
\end{equation}
arising from the argument of the square root in (\ref{definition_LBI}). 

Inspection of (\ref{BI_wave_parameters}) shows that the phase velocity $v$ of the wave is equal to or less than the speed of light $c=1$ in vacuum Maxwell theory. Thus, it is possible for a particle to propagate faster than the phase speed $v$ of the wave described by (\ref{exact_wave_BI}). Although a perturbative analysis in $\kappa$ would simply lead to corrections to predictions of linear Maxwell theory, we expect Born-Infeld electrodynamics to have novel implications resulting from non-perturbative considerations. Indeed, as we will now show, a test electron driven by the wave (\ref{exact_wave_BI}) can have constant $4$-acceleration and this result has no analogue in vacuum Maxwell electrodynamics.

As in Section~\ref{sec:non-lin_plasma_wave}, the worldline $C : \tau \mapsto (t(\tau), x(\tau), y(\tau), z(\tau))$ of the test electron is presumed to satisfy the Lorentz equation
\begin{equation}
\label{test_lorentz_rpt}
m\nabla_{\dot{C}}\dot{C} = q\widetilde{\iota_{\dot{C}}F}
\end{equation}
and the normalization condition
\begin{equation}
\label{test_norm}
g(\dot{C},\dot{C})=-1
\end{equation}
where $\dot{C} = \dot{t}\, \partial/\partial t +  \dot{x}\, \partial/\partial x +  \dot{y}\, \partial/\partial y +  \dot{z}\, \partial/\partial z$ with $\dot{t} = dt/d\tau$,  $\dot{x} = dx/d\tau$,  $\dot{y} = dy/d\tau$.  $\dot{z} = dz/d\tau$.

The properties of the Born-Infeld plane wave solution (\ref{exact_wave_BI}) suggest that it may be fruitful to ask whether solutions to (\ref{test_lorentz_rpt}), (\ref{test_norm}) exist that have constant phase, i.e. $\dot{z} - v \dot{t} = 0$. This condition may be written as $\dot{C}\zeta = 0$ where $\zeta=z-vt$ as before, and an electron satisfying $\dot{C}\zeta = 0$ may be envisaged as `surfing' the wave. However, although we are free to choose $\dot{C}\zeta |_{\tau=0} = 0$, there is no  guarantee that a solution to (\ref{test_lorentz_rpt}), (\ref{test_norm}) exists such that $\dot{C}\zeta$ vanishes at $\tau > 0$. However, as we will now show, such solutions do exist if the properties of the profile ${\cal E}$ are appropriate.

In the following, it is convenient to introduce the type-$(1,1)$ tensor ${\cal F}$ that satisfies ${\cal F}(\alpha,U) = \alpha(\widetilde{\iota_U F})$ for all choices of $1$-form $\alpha$ and vector $U$. Furthermore, to avoid an unnecessary plethora of brackets or indices, it is useful to denote tensor contraction by juxtaposition; in particular, $\alpha U = \alpha (U)$, $\alpha{\cal F}U = {\cal F}(\alpha,U)$ and ${\cal F}{\cal G} = {\cal F}(-,\partial_a) \otimes {\cal G}(dx^a,-)$ where ${\cal G}$ is a type-$(1,1)$ tensor and the Einstein summation convention is used with $\{d x^a\} = \{dt, dx, dy, dz\}$, $\{\partial_a\} = \{\partial/\partial t, \partial/\partial x,\partial/\partial y,\partial/\partial z\}$ and $a = 0,1,2,3$.

The constraint $\dot{C}\zeta = 0$ may be written as
\begin{equation}
d\zeta \dot{C} = 0
\label{first_constraint}
\end{equation}
and a consequence of (\ref{first_constraint}) is that its derivative $\nabla_{\dot{C}}(d\zeta\dot{C})$ along $C$ must vanish. Thus, a second constraint
\begin{equation}
d\zeta {\cal F} \dot{C} = 0
\label{second_constraint}
\end{equation}
is generated from the first constraint (\ref{first_constraint}) using (\ref{test_lorentz_rpt}) and $\nabla_{\dot{C}} d\zeta = 0$.

Since inspection of (\ref{exact_wave_BI}) reveals that $\nabla_{\dot{C}} {\cal F}$ is proportional to $d\zeta\dot{C}$, it follows that $\nabla_{\dot{C}} {\cal F}$ vanishes as a consequence of (\ref{first_constraint}). Hence, differentiation of the second constraint (\ref{second_constraint}) along $C$ yields the third constraint
\begin{equation}
d\zeta {\cal F}^2 \dot{C} = 0
\label{third_constraint}
\end{equation}
and, likewise, differentiation of (\ref{third_constraint}) along $C$ leads to the fourth constraint
\begin{equation}
d\zeta {\cal F}^3 \dot{C}=0
\label{fourth_constraint}
\end{equation}
and so-on, where ${\cal F}^n \equiv \Pi^n_{p=1} {\cal F}$. However, the Cayley-Hamilton theorem can be invoked to write ${\cal F}^n$, for $n \ge 4$, in terms of a linear superposition of ${\cal F}^3$, ${\cal F}^2$, ${\cal F}$, and the identity tensor, so no new constraints are generated for $n>3$.

A condition on the electromagnetic field is obtained when (\ref{first_constraint}), (\ref{second_constraint}), (\ref{third_constraint}), (\ref{fourth_constraint}) are regarded as a linear system for the components of $\dot{C}$. Then, a non-trivial $\dot{C}$ exists if and only if $\Lambda = 0$ where the $4$-form $\Lambda$ is
\begin{equation}
\Lambda = d\zeta \wedge d\zeta{\cal F} \wedge d\zeta{\cal F}^2 \wedge d\zeta{\cal F}^3.
\label{det_condition}
\end{equation}
It can be shown that the $0$-form $\star \Lambda$ is a cubic polynomial in ${\cal E}$ with one simple root ${\cal E}=0$ and one repeated root ${\cal E}={\cal E}_{\rm crit}$ where the critical value ${\cal E}_{\rm crit}$ of the wave profile ${\cal E}$ is the remarkably simple expression
\begin{equation}
{\cal E}_{\rm crit} = \frac{1 + \kappa^2 B^2}{\kappa^2 B_y}.
\label{E_crit}
\end{equation}

The type-$(1,1)$ tensor ${\cal F}_{\rm crit} = {\cal F}|_{{\cal E} = {\cal E}_{\rm crit}}$ has a number of interesting properties. Firstly, $d\zeta$ is an eigenform of ${\cal F}_{\rm crit}^2$, and it follows that (\ref{test_norm}), (\ref{first_constraint}), (\ref{second_constraint}) are the only independent algebraic conditions on $\dot{C}$ that arise from the above analysis. Secondly, the bound (\ref{XY_bound}) is saturated by ${\cal F} = {\cal F}_{\rm crit}$, and the significance of this result is revealed by recalling that the term within the square root in (\ref{definition_LBI}) may be written as a determinant :
\begin{equation}
1 - \kappa^2 X - \kappa^4 Y^2/4 = {\rm det}(I + \kappa{\cal F})
\label{poly_det_BI}
\end{equation} 
where $I$ is the type-$(1,1)$ identity tensor. Thus,
\begin{equation}
{\rm det}(I + \kappa{\cal F}_{\rm crit}) = 0
\end{equation}
and, hence, $-1/\kappa$ is an eigenvalue of ${\cal F}_{\rm crit}$. Morever, ${\rm det}(I + \kappa{\cal F}) = {\rm det}(I - \kappa{\cal F})$ because ${\cal F}$ is generated from a totally antisymmetric tensor (the $2$-form $F$), and it follows that $1/\kappa$ is also an eigenvalue of ${\cal F}_{\rm crit}$. Furthermore, the eigenvectors $W_+$, $W_-$ satisfying
\begin{align}
\label{W_+_def}
&{\cal F}_{\rm crit} W_+ = \frac{1}{\kappa} W_+,\\
\label{W_-_def}
&{\cal F}_{\rm crit} W_- = -\frac{1}{\kappa} W_-,
\end{align}
are null with respect to the spacetime metric,
\begin{equation}
\label{lightlike_W}
\widetilde{W_+} W_+ = 0,\qquad \widetilde{W_-} W_- = 0
\end{equation} 
which follows because $\kappa \widetilde{W_+} {\cal F}_{\rm crit} W_+ = \widetilde{W_+}W_+$, $\kappa \widetilde{W_-} {\cal F}_{\rm crit} W_- = -\widetilde{W_-}W_-$ and $\widetilde{U}{\cal F} U = \iota_U \iota_U F = 0$ for all $U$. Finally, using (\ref{exact_wave_BI}), (\ref{BI_wave_parameters}), (\ref{E_crit}) it can be shown that
\begin{equation}
d\zeta {\cal F}^2_{\rm crit} = \lambda\,d\zeta
\end{equation}
where
\begin{equation}
\lambda = - \frac{B_x^2 (1+\kappa^2 B^2)}{\kappa^2 B_y^2 (1 + \kappa^2 B_z^2)}.
\end{equation} 
However, (\ref{W_+_def}), (\ref{W_-_def}) yield $d\zeta {\cal F}^2_{\rm crit} W_+ = d\zeta W_+ / \kappa^2$, $d\zeta {\cal F}^2_{\rm crit} W_- = d\zeta W_- / \kappa^2$, respectively, and therefore
\begin{align}
\label{dzeta_W}
&d\zeta W_+ =0,\qquad d\zeta W_- = 0
\end{align}
since $\lambda \neq 1/\kappa^2$.

The above considerations demonstrate that the pair $\{W_+, W_-\}$ is a natural basis for constructing solutions to (\ref{test_lorentz_rpt}), (\ref{test_norm}), (\ref{first_constraint}), (\ref{second_constraint}).  We can choose $\widetilde{W_+} W_- = -1/2$ and choose $W_+, W_-$ to be future directed without loss of generality and, using the above results, it is easy to see that
\begin{equation}
\label{dot_C}
\dot{C} = \exp\bigg(\frac{q}{m\kappa}\tau\bigg) W_+  + \exp\bigg(-\frac{q}{m\kappa}\tau\bigg) W_-
\end{equation}
satisfies (\ref{test_norm}), (\ref{first_constraint}), (\ref{second_constraint}). Furthermore, (\ref{test_lorentz_rpt}) is satisfied because ${\cal \nabla}_{\dot{C}} W_+ = 0$, ${\cal \nabla}_{\dot{C}} W_- = 0$. The non-perturbative nature of (\ref{dot_C}) in $\kappa$ is clearly visible in the argument of the exponentials. 

The asymptotic behaviour, as $\tau \rightarrow \pm \infty$, of a test electron with $4$-velocity (\ref{dot_C}) is determined by the pair $\{W_+, W_-\}$. The $3$-velocity $\bm{u} = (u_x, u_y, u_z)$ of the electron is related to the $4$-velocity $\dot{C}$ as
\begin{equation}
\dot{C} = \frac{1}{\sqrt{1-\bm{u}^2}}\bigg(\frac{\partial}{\partial t} + u_x \frac{\partial}{\partial x} + u_y \frac{\partial}{\partial y} + u_z \frac{\partial}{\partial z}\bigg)   
\end{equation}
where $\bm{u}^2 = u_x^2 + u_y^2 + u_z^2$. Hence, recalling $q<0$, it follows that
\begin{align}
\underset{\tau \rightarrow -\infty}{\lim} \bm{u} = \frac{d\bm{x} W_+}{dt W_+}, \qquad \underset{\tau \rightarrow \infty}{\lim} \bm{u} = \frac{d\bm{x} W_-}{dt W_-}.
\end{align}
where $d\bm{x}W_+ = (dxW_+,\, dyW_+,\, dzW_+)$, $d\bm{x}W_- = (dxW_-,\, dyW_-,\, dzW_-)$.
Inspection of (\ref{dzeta_W}), (\ref{lightlike_W}) shows that, without loss of generality, the $4$-vectors $W_+/dt W_+$ and $W_-/dt W_-$ may be parameterised as
\begin{align}
\label{W_+_alpha}
&\frac{W_+}{dt W_+} = \frac{\partial}{\partial t} + v \frac{\partial}{\partial z} + \sqrt{1-v^2}\bigg(\cos\alpha_+ \frac{\partial}{\partial x} +   \sin\alpha_+ \frac{\partial}{\partial y}\bigg),\\
\label{W_-_alpha}
&\frac{W_-}{dt W_-} = \frac{\partial}{\partial t} + v \frac{\partial}{\partial z} + \sqrt{1-v^2}\bigg(\cos\alpha_- \frac{\partial}{\partial x} +   \sin\alpha_- \frac{\partial}{\partial y}\bigg)
\end{align}
where $\alpha_+, \alpha_-$ are the angles (from the $x$-axis) of the incoming and outgoing trajectories, respectively, projected into the $x-y$ plane. Consideration of (\ref{W_+_def}), (\ref{W_-_def}), (\ref{W_+_alpha}), (\ref{W_-_alpha}) yields
\begin{align}
\label{sin_cos_beta}
\sin(\alpha_\pm - \varphi) = \pm \frac{1}{\sqrt{1 + \kappa^2 B^2 \cos^2\theta}},\qquad \cos(\alpha_\pm - \varphi) = \frac{\kappa B \cos\theta}{\sqrt{1 + \kappa^2 B^2 \cos^2\theta}}
\end{align}
where $\theta, \varphi$ are spherical polar angles that specify the orientation of the ambient magnetic field relative to the $z$-axis:
\begin{align}
B_x = B \sin\theta\cos\varphi, \quad B_y = B \sin\theta\sin\varphi,\quad B_z = B \cos\theta.
\end{align}
The above results may be used to qualitatively estimate the behaviour of an electron captured in the peak of a pulse propagating at speed $v$. We see that the electron emanates from the core of the pulse with $3$-velocity $\bm{u}_- = (\sqrt{1-v^2} \cos \alpha_-, \sqrt{1-v^2} \sin \alpha_-, v)$ where $\alpha_- = \varphi + \text{arccot}(-\kappa B \cos\theta)$. Hence, for physically reasonable values of $\kappa B$, the electron is strongly accelerated to near the speed $c=1$ and ejected from the pulse at a finite angle to the direction of propagation of the pulse. In particular, $\bm{u}_- = (-\kappa B \sin\theta \sin\varphi, \kappa B\sin\theta \cos\varphi, 1) + {\cal O}(\kappa^2 B^2)$ and the electron is ejected from the core of the pulse at the azimuthal angle $\varphi + \pi/2 +{\cal O}(\kappa B)$.
%%%
\section{Conclusion}
%%%
Two distinct non-perturbative results have been presented that address particle acceleration in non-linear electrodynamics. In the absence of an established theory of radiation reaction in the context of non-linear electrodynamics, we focussed our attention on a simple matter model compatible with stress-energy-momentum balance and explored test particle motion in that context. In Section~\ref{sec:non-lin_plasma_wave} we obtained an expression for the maximum energy gained by a test electron in a non-linear density wave in a magnetised plasma. The expression is valid for a wide range of physically permissible theories of non-linear electrodynamics encoded by a Lagrangian of the two electromagnetic invariants (including Born-Infeld electrodynamics). However, if Born-Infeld electrodynamics emerges at sufficiently strong field intensities then, as shown in Section~\ref{sec:non-lin_vacuum_wave}, an outcome may occur that cannot be realised using perturbative considerations based the linear vacuum Maxwell equations. Under the appropriate conditions, a test electron can `surf' a critically intense Born-Infeld electromagnetic plane wave and be strongly accelerated by the wave.
\section{Acknowledgments}
%%%%
This work was undertaken as part of the ALPHA-X consortium funded under EPSRC grant EP/J018171/1 and with support from the Cockcroft Institute of Accelerator Science and Technology  (STFC grant ST/G008248/1).

\end{document}